\documentclass[fleqn,12pt]{article}
\usepackage{espcrc2}

\usepackage{graphicx}
\usepackage[figuresright]{rotating}

\newcommand{\AmS}{{\protect\the\textfont2
  A\kern-.1667em\lower.5ex\hbox{M}\kern-.125emS}}

\title{How the recent BABAR data for $P\to \gamma\gamma^*$ affect the Standard Model predictions for the rare decays $P\to l^+l^-$.}

\author{A.E. Dorokhov\address[MCSD]{Joint Institute for Nuclear Research,
Bogoliubov Laboratory of Theoretical
Physics, \\  141980 Dubna, Moscow region, Russian Federation; \\
Institute for Theoretical Problems of Microphysics, Moscow State University, \\  RU-119899, Moscow, Russian Federation}%
        \thanks{The author acknowledges partial support from the Scientific School grant 3159.2010.2.}}

\begin{document}

\begin{abstract}
Measuring the lepton anomalous magnetic moments $(g-2)$ and the rare decays of light pseudoscalar mesons into lepton pairs $P\rightarrow l^{+}l^{-} $, serve as important tests of the Standard Model. To reduce the theoretical uncertainty in the standard model predictions, the data on the charge and transition form factors of the light pseudoscalar mesons play a significant role. Recently, new data on the behavior of the transition form factors $P\to\gamma\gamma^*$ at large momentum transfer were supplied by the BABAR collaboration. There are several problems with the theoretical interpretation of these data: 1) An unexpectedly slow decrease of the pion transition form factor at high momenta, 2) the qualitative difference in the behavior of the pion form factor and the $\eta$ and $\eta^\prime$ form factors at high momenta, 3) the inconsistency of the measured ratio
of the $\eta$ and $\eta^\prime$ form factors with the predicted one. We comment on the influence of the new BABAR data on the rare decay branchings.

\vspace{1pc}
\end{abstract}

% typeset front matter (including abstract)
\maketitle

Modern cosmology tells us that $95\%$ of the matter in the universe is not described in terms of the Standard Model (SM) matter. New excitement appeared after the satellite experiments Fermi LAT, PAMELA, ATIC, HESS and WMAP that provided data which give an indication interpreted as Dark Matter and/or Pulsar signals. Thus the search for hints of Physics beyond SM is the fundamental problem of modern elementary particle physics. There are two strategies to solve this problem.

Firstly, in high energy experiments an enormous amount of energy is applied in order to excite the heavy degrees of freedom expected to be associated with the new physics. At the moment, there is not any evidence on a deviation between the SM predictions and the high energy data, and we are urgently waiting for the results of the physical program at LHC.

Another strategy is to carry out low energy experiments, where the lack of energy is compensated by a huge statistics producing rare processes related to the new interactions. The low energy experiments are not only supplement to the high energy program, allowing to get combined restrictions on the parameters of hypothetical interactions, but they are also unique instrument for the discovery of the physics beyond the SM, containing low mass particles. At the moment, there are some problems with the matching of the experimental data with the predictions of the SM. The most famous one is the descrepancy by three standard deviations between experiment \cite{Bennett:2006fi} and SM theory \cite{Davier:2009zi} for the muon anomalous magnetic moment (AMM). Another example is similarly large deviation between the recent precise experimental result on the rare $\pi^0$ decay into $e^-e^+$ pair \cite{Abouzaid:2006kk} and the SM prediction \cite{Dorokhov:2007bd,Dorokhov:2008cd,Dorokhov:2008qn,Dorokhov:2009xs}.

At the early stage of the study of the lepton anomalous magnetic moments, $a_l=(g_l-2)/2$, entering the structure of the vector vertex
\begin{equation}
\Gamma_\mu=e\gamma_\mu+a_l\frac{ie}{2m_l}\sigma_{\mu\nu}q_\nu,
\label{AMM}
\end{equation}
played a fundamental role in the foundations of quantum mechanics and, in particular, of quantum field theory \cite{Schwinger:1948iu}. At present, the accent of the study is shifted to the test of the SM and the search of the physics beyond it. Within the SM, the dominant contribution to the lepton AMM is due to the QED, supplemented by small, but visible corrections from the strong and weak interactions.

The electron AMM is measured with one of the best accuracies obtained for physical observables \cite{Hanneke:2008tm}
\begin{equation}
a_e^{\mathrm{exp}}=1 159 652 180. 73 (0.28)\cdot 10^{-12} [0.28 \mathrm{ppt}].
\label{EAMM}
\end{equation}
In the SM, it is given by
\begin{eqnarray}
a_e^{\mathrm{SM}}=a_e^{\mathrm{QED}}+a_e^{\mathrm{hadron}}+a_e^{\mathrm{weak}},\nonumber\\
a_e^{\mathrm{QED}}=\sum_{n=1}^5C_{2n}\left(\frac{\alpha}{\pi}\right)^n+...,
\label{EAMM_SM}
\end{eqnarray}
where the first three coefficients are known analytically and the two others in some approximations.  If the fine structure constant $\alpha$ would be known from other independent sources, the measurement of the electron AMM would be a stringent test of QED. However, the theoretical error is dominated by the
uncertainty in the input value for the QED coupling constant $\alpha$, and the problem is reversed to that of obtaining a best estimate of the QED coupling constant \cite{Hanneke:2008tm}
\begin{equation}
\alpha^{-1}= 137.035 999 084 (51) [0.37 \mathrm{ ppb}].
\label{alpha}
\end{equation}

The great feature of the study of the leptonic AMM is, that they are very sensitive to the manifestation of new physics. Any interaction with characteristic scale $\Lambda$ contributes to the leptonic AMM like\footnote{An alternative mechanism for a contribution by new physics was proposed in \cite{Chizhov:2008yt}. It occurs at the tree level and exhibits a linear rather than quadratic dependence on $m_l$.}: $(m_l/\Lambda)^2$. Therefore, the heavier the lepton, the more visible the interaction. In this way, the contribution of an interaction to the muon AMM is bigger than that to the electron AMM by a factor $(m_\mu/m_e)^2\approx 10^4$. Even bigger would be the effect for the $\tau$-lepton AMM. However, the $\tau$-lepton is highly unstable, and the measurement of its AMM is problematic (see \cite{Eidelman:2007sb} for discussions). At moment, there are only very rough experimental limitations on the $\tau$-lepton AMM set by the  L3 \cite{Acciarri:1998iv}, OPAL \cite{Ackerstaff:1998mt} and DELPHI collaborations \cite{Abdallah:2003xd,Boiko:2005ws},
 \begin{eqnarray}
-0.052 < a_\tau < 0.058,\ \ \ \ \ \ \ \mathrm{L3},\nonumber\\
-0.068 < a_\tau < 0.065,\ \ \ \ \ \ \ \mathrm{OPAL},\nonumber\\
-0.052 < a_\tau < 0.013,\ \ \ \ \ \ \ \mathrm{DELPHI}
\label{TauAMM}
\end{eqnarray}
from $Z\to\tau\tau\gamma$ and $e^+e^-\to e^+e^-\tau^+\tau^-$ processes, while the SM prediction is \cite{Eidelman:2007sb}
\begin{equation}
a_\tau^{\mathrm{SM}}=1. 17721 (5)\cdot 10^{-3}.
\label{TauAMM SM}
\end{equation}

The theoretical studies of the muon AMM $g-2$ (see for review \cite{Miller:2007kk,Passera:2007fk,Dorokhov:2005ff,Jegerlehner:2009ry,Prades:2009qp}), the rare decays of light pseudoscalar mesons  into lepton pairs \cite{Dorokhov:2007bd,Dorokhov:2008cd,Dorokhov:2008qn,Dorokhov:2009xs} and the comparison with the experimental results, offer an important low-energy tests of the SM. The discrepancy between the present SM prediction of the muon AMM and its experimental determination \cite{Bennett:2006fi} is $(24.6\pm 8.0)\cdot 10^{-10}$ ($3.1\sigma$) \cite{Prades:2009qp}. The situation with the rare decays of the light pseudoscalar mesons into lepton pairs became more intriguing after the recent KTeV E799-II experiment at FermiLab \cite{Abouzaid:2006kk} in which the pion decay into an electron-positron pair was measured with high accuracy using the $K_{L}\rightarrow3\pi$ process as a source of tagged neutral pions ($R\left(  P\rightarrow l^{+}l^{-}\right)=\Gamma\left(  P\rightarrow l^{+}l^{-}\right)/\Gamma_{tot}$)

\begin{equation}
R^{\mathrm{KTeV}}\left( \pi^{0}\rightarrow e^{+} e^{-} \right)
 =\left( 7.49\pm0.38 \right) \cdot10^{-8}.
\label{KTeV}
\end{equation}
The standard model prediction gives \cite{Dorokhov:2007bd,Dorokhov:2009xs}
\begin{equation}
 R^{\mathrm{Theor}}\left(  \pi^{0}\rightarrow e^{+}e^{-}\right)
 =\left( 6.2\pm0.1\right)  \cdot10^{-8},
\label{Bth}
\end{equation}
which is $3.1\sigma$ below the KTeV result (\ref{KTeV}).

The main limitation on realistic predictions for these processes originates from the large distance contributions of the strong sector of the SM, where perturbative QCD does not work. In order to diminish the theoretical uncertainties, the use of the experimental data on the pion charge and transition form factors are of crucial importance. The first one, measured in $e^+e^- \to \pi^+\pi^-(\gamma)$ by CMD-2 \cite{Akhmetshin:2006bx}, SND \cite{Achasov:2006vp}, KLOE \cite{Aloisio:2004bu}, and BABAR \cite{Aubert:2009fg} provides an estimate for the hadron vacuum polarization contribution to muon $g-2$, with accuracy better than $1\%$. The second one, measured in $e^+e^- \to e^+e^-P$ for spacelike photons by CELLO \cite{Behrend:1990sr}, CLEO \cite{Gronberg:1997fj}, and BABAR \cite{:2009mc} collaborations and in $e^+e^- \to P\gamma$ for timelike photons by the BABAR \cite{Aubert:2006cy} collaboration, is essential to reduce the theoretical uncertainties in the estimates of the contributions of the hadronic light-by-light process to the muon $g-2$ and in the estimates of the decay widths of $P\to l^+l^-$. The BABAR data \cite{Aubert:2006cy,:2009mc} on the large momentum behavior of the form factors cause the following problems for their theoretical interpretation: 1) An unexpectedly slow decrease of the pion transition form factor at high momenta \cite{:2009mc}, 2) the qualitative difference in the behavior of the pion and $\eta,\eta^\prime$ form factors at high momenta \cite{Dorokhov:2009zx}, 3) inconsistency of the measured ratio of the $\eta,\eta^\prime$ form factors with the predicted one \cite{Aubert:2006cy}.

In Figs. 1-3 the data for the $\pi^0$, $\eta$ and $\eta'$ transition form factors from the CELLO, CLEO, and BABAR collaborations are presented. In Figs. 2 and 3, the BABAR point, measured in the timelike region \cite{Aubert:2006cy}, is drawn at $Q^2=112$ GeV$^2$, assuming that the spacelike and timelike asymptotics of the form factor are equal. It is seen from the Figs. 2 and 3, that the spacelike and timelike points are conjugated. We hope that new data from the BABAR and BELLE collaborations confirm this assumption.

%%%%%%%%%%%%%%%%%%%%%%%%%%%%%%%%%FIGURE 1%%%%%%%%%%%%%%%%%%%%%%%%%%%%%%

\begin{figure}[ht]
\hspace*{-10mm}\includegraphics[width=0.6\textwidth]{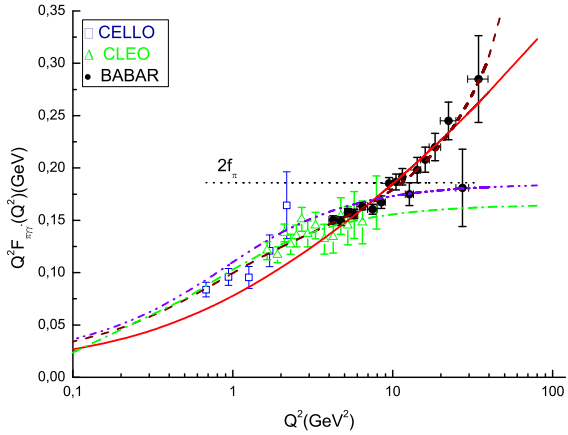}  \vspace*{-10mm}
\caption{{\protect\footnotesize The transition form factor $\protect\gamma^\ast\protect\gamma\rightarrow \protect\pi
^{0}$. The data are from the CELLO
\protect\cite{Behrend:1990sr}, CLEO \protect\cite{Gronberg:1997fj} and BABAR
\protect\cite{:2009mc} Collaborations. The dotted line is massless QCD
asymptotic limit. (The notation for curves is explained in the text.)}}
\label{fig:pi}
\end{figure}
%%%%%%%%%%%%%%%%%%%%%%%%%%%%%%%%%%%%%%%%%%%%%%%%%%%%%%%%%%%%%%%%%%%%%%%

%%%%%%%%%%%%%%%%%%%%%%%%%%%%%%%%%FIGURE 1%%%%%%%%%%%%%%%%%%%%%%%%%%%%%%
\begin{figure}[ht]
\hspace*{-10mm}\includegraphics[width=0.6\textwidth]{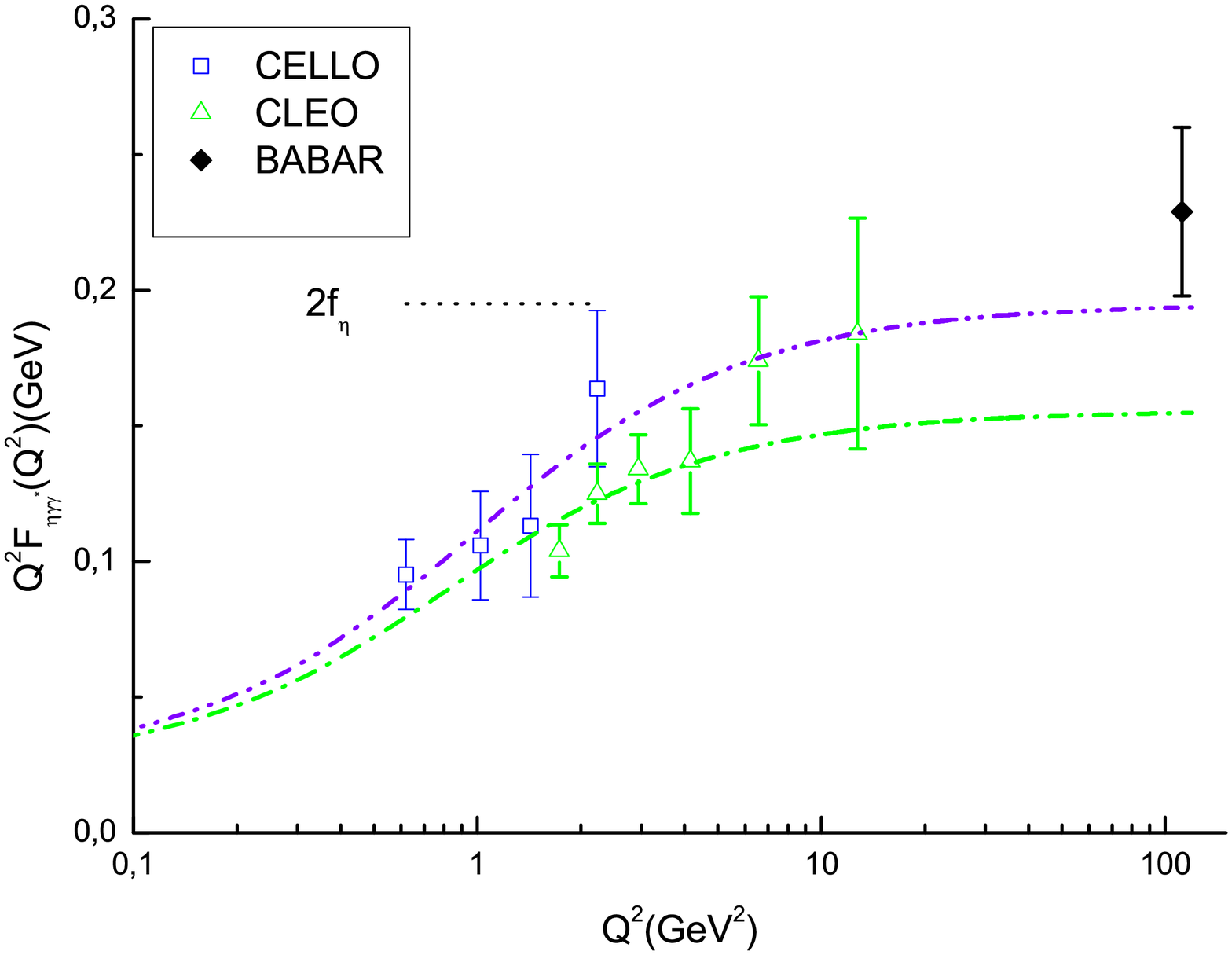}  \vspace*{-10mm}
\caption{{\protect\footnotesize The transition form factor $\protect\gamma^\ast\protect\gamma\rightarrow \protect\eta$.
The data are from the CELLO \protect\cite{Behrend:1990sr}, and CLEO \protect\cite{Gronberg:1997fj} Collaborations.
The CLEO results obtained in different $\eta$ decay modes are averaged.
The BABAR point, measured in the timelike region $\protect\gamma^\ast\rightarrow \protect\eta\protect\gamma$ \cite{Aubert:2006cy}, is drawn at $Q^2=112$ GeV$^2$, assuming that the spacelike and timelike asymptotics of the form factor are equal.
(The notation for curves is explained in the text.)}}
\label{fig:eta}
\end{figure}
%%%%%%%%%%%%%%%%%%%%%%%%%%%%%%%%%%%%%%%%%%%%%%%%%%%%%%%%%%%%%%%%%%%%%%%

%%%%%%%%%%%%%%%%%%%%%%%%%%%%%%%%%FIGURE 1%%%%%%%%%%%%%%%%%%%%%%%%%%%%%%
\begin{figure}[ht]
\hspace*{-10mm}\includegraphics[width=0.6\textwidth]{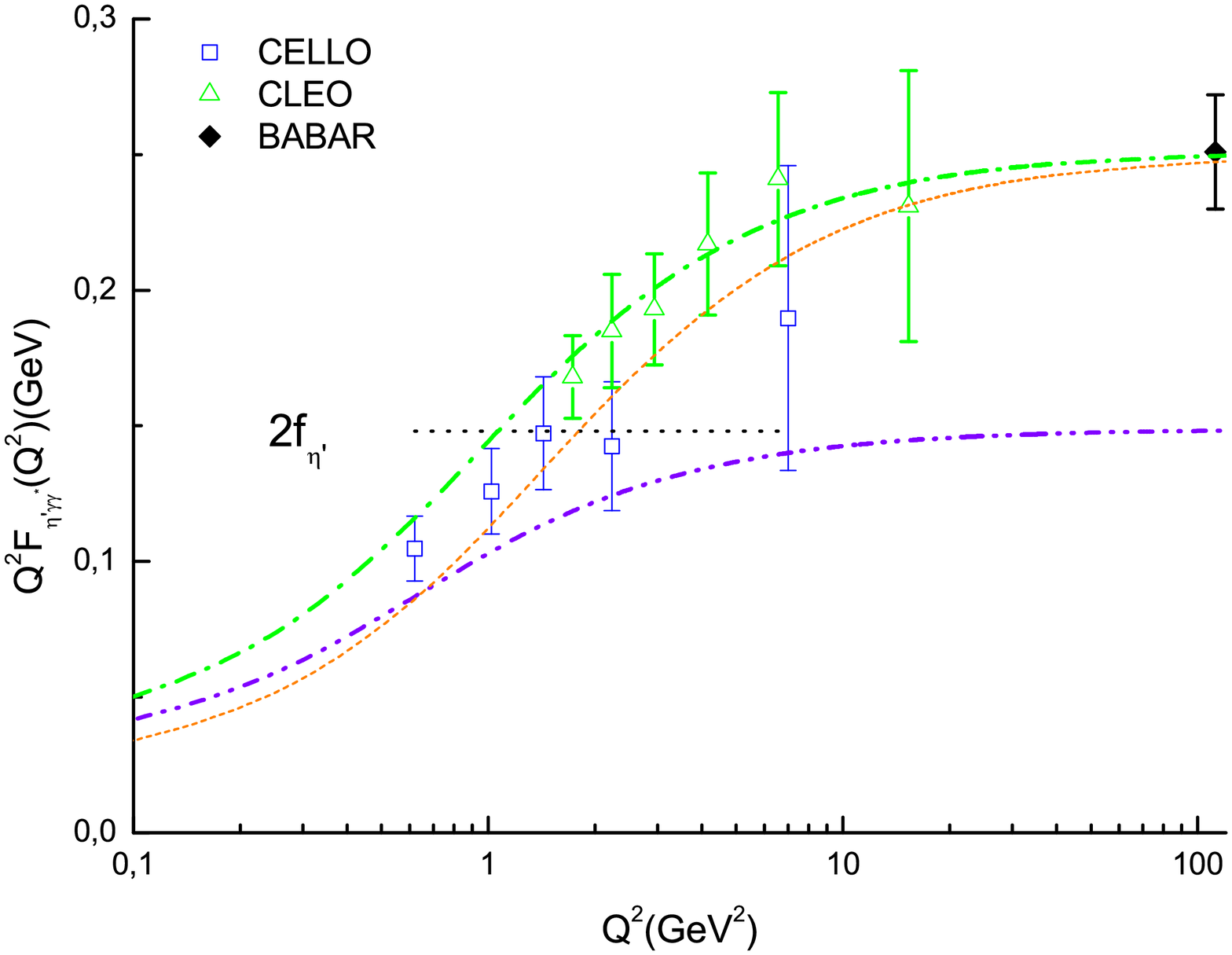}  \vspace*{-10mm}
\caption{{\protect\footnotesize The transition form factor $\protect\gamma^\ast\protect\gamma\rightarrow \protect\eta'$.
The data are from the CELLO \protect\cite{Behrend:1990sr}, and CLEO \protect\cite{Gronberg:1997fj}
\protect\cite{Aubert:2006cy} Collaborations. The CLEO results obtained in different $\eta'$ decay modes are averaged.
The BABAR point, measured in the timelike region $\protect\gamma^\ast\rightarrow \protect\eta'\protect\gamma$ \cite{Aubert:2006cy}, is drawn at $Q^2=112$ GeV$^2$, assuming that the spacelike and timelike asymptotics of the form factor are equal.
The dashed line is the perturbative QCD asymptotic limit. (The notation for curves is explained in the text.)}}
\label{fig:eta1}
\end{figure}
%%%%%%%%%%%%%%%%%%%%%%%%%%%%%%%%%%%%%%%%%%%%%%%%%%%%%%%%%%%%%%%%%%%%%%%

At zero momentum transfer, the transition form factor is fixed by the two-photon decay width
\begin{equation}
F_{P\gamma\gamma^*}^2(0,0)=\frac{1}{(4\pi\alpha)^2}\frac{64\pi\Gamma(P\to\gamma\gamma)}{M_P^3},
\label{F0exp}\end{equation}
where $\alpha$ is the QED coupling constant, $M_P$ is the resonance mass and
$\Gamma(P\to\gamma\gamma)$ is the two-photon partial width of the meson $P$. The axial anomaly predicts
\begin{equation}
F_{P\gamma\gamma^*}(0,0)\approx\frac{1}{4\pi^2f_P},
\label{F0the}\end{equation}
where $f_P$ is the meson decay constant. Under assumption of factorization, perturbative QCD predicts the asymptotic behavior of the $F_{P\gamma\gamma^*}^2(Q^2,0)$ transition form factors as $Q^2\to\infty$ \cite{Brodsky:1981rp}
\begin{equation}
F_{P\gamma\gamma^*}(Q^2\to\infty,0)\sim\frac{2f_P}{Q^2}.
\label{Fas}\end{equation}
The perturbative QCD corrections to this expression at large momentum transfer are extremely small \cite{Chase:1979ck,Braaten:1982yp,Kadantseva:1985kb,Mikhailov:2009kf}.

To describe the soft nonperturbative region of $Q^2$, a simple interpolation between $Q^2\to0$ and $Q^2\to\infty$ limits has been proposed by Brodsky and Lepage (BL) \cite{Brodsky:1981rp}:
\begin{eqnarray}
&&F^{\rm{BL}}_{\pi\gamma\gamma^*}(Q^2,0)=\frac{1}{4\pi^2f_P}\frac{1}{1+Q^2/(\Lambda^{BL}_P)^2},\nonumber\\
&&(\Lambda^{BL}_P)^2=8\pi^2f^2_P, \label{Fbl}
\end{eqnarray}
where the values of $f_P$ are estimated from (\ref{F0exp}) and  (\ref{F0the}) \cite{Gronberg:1997fj}: $f_\pi=92.3$ MeV, $f_\eta=97.5$ MeV, $f_{\eta'}=74.4$ MeV.

The CLEO (and CELLO) collaboration parameterized their data by a formula similar to (\ref{Fbl}), but with the pole mass being a free fitting parameter \cite{Gronberg:1997fj},
\begin{equation}
F^{\rm{CLEO}}_{\pi\gamma\gamma^*}(Q^2,0)=\frac{1}{4\pi^2f_P}\frac{1}{1+Q^2/\Lambda_P^2},
\label{Fcleo}\end{equation}
where $\Lambda_\pi=776\pm22$ MeV, $\Lambda_\eta=774\pm29$ MeV, and $\Lambda_{\eta'}=859\pm28$ MeV.

In Figs. 1-3 the asymptotics (\ref{Fas}) are shown by dotted lines, the BL interpolations (\ref{Fbl}) are given by dot-dot-dashed lines, and the CLEO parametrizations extrapolated to higher momentum transfer  (\ref{Fcleo}) are shown by dot-dashed lines. We see that the QCD inspired expression (\ref{Fbl}) works well only for the $\eta$ meson form factor with $\Lambda^{BL}_\eta=\sqrt{8\pi^2f^2_\eta}\approx866$ MeV (Fig. 2), whereas the CLEO parametrization (\ref{Fcleo}) underestimates the large $Q^2$ behavior. On the other hand, the CLEO parametrization describes the $\eta'$ meson form factor (Fig. 3) well, but the BL expression strongly underestimates the large $Q^2$ behavior. We still have a good description of the $\eta'$ meson form factor by BL formula if one takes $f_{\eta'}=125$ MeV (short dashed line in Fig. 3), but then the normalization is incorrect.

For the $\eta$ and $\eta'$ mesons the parametrizations (\ref{Fbl}) and (\ref{Fcleo}) correctly reflect the experimental data at large $Q^2$ on the qualitative level. This is not the case for the pion form factor showing the growth at large  $Q^2$, which is unexpected from the QCD factorization approach \cite{Mikhailov:2009kf} (Fig. 1). However, this growth is easy to explain \cite{Dorokhov:2009dg}  in the context of the quark model \cite{Quark2}. Within this model, the pion form factor is given by the quark-loop (triangle) diagram with momentum independent quark mass serving as an infrared regulator \cite{Pivovarov:2001mw}. The form factor has double logarithmic asymptotics at large $Q^2$: $\ln^2(Q^2/M_q^2)$ and is given by \cite{Quark2}
\begin{eqnarray}
&&F_{\pi\gamma\gamma^*}(Q^2,0)=\frac{m_\pi^2}{m_\pi^2+Q^2}\frac{1}{%
2\arcsin^2(\frac{m_\pi}{2M_Q})}  \nonumber \\
&&\cdot\{2\arcsin^2(\frac{m_\pi}{2M_Q})+\frac{1}{2}\ln^2\frac{\beta_Q+1}{%
\beta_Q-1}\}.  \label{Ftt}
\end{eqnarray}
where $\beta_Q=\sqrt{1+\frac{4M_Q^2}{Q^2}}$. The solid line in Fig. 1 is the pion transition form factor calculated from Eq. (\protect\ref{Ftt}) with the parameter $M_Q=135$ MeV. The advantage of this model is, that it defines the form factor for arbitrary virtuality of photons and has correct normalization at zero virtualities. One of disadvantages is, that the corresponding integral for the decay constant $f_\pi$ is divergent and should be regularized. Another feature of the model is, that the vertices and propagators are local. In particular, the pion-to-quarks vertex is local, just like in the Nambu-Jona-Lasinio model. It is known, that in this case the pion distribution amplitude and distribution function are constants \cite{Davidson:1994uv,RuizArriola:2002bp,Dorokhov:2000gu,Anikin:1999cx}.

The flat (almost constant) pion distribution amplitude became popular in the context of the explanation of the BABAR data within different factorization schemes \cite{Radyushkin:2009zg,Polyakov:2009je,Li:2009pr,Kochelev:2009nz}. For example, in the model \cite{Radyushkin:2009zg} the pion transition form factor is
\begin{eqnarray}
&&F_{\pi\gamma\gamma^*}(Q^2,0)=\frac{2}{3}\frac{f_\pi}{Q^2}  \nonumber \\
&&\cdot\int_0^1\frac{dx}{x}\left[1-\exp{\left(-\frac{xQ^2}{2\sigma (1-x)}%
\right)}\right].  \label{Frad}
\end{eqnarray}
and has logarithmically enhanced asymptotic behavior $\sim \log{\left(1+Q^2/\sigma\right)}$. In the kinematical range of Fig. 1, the model (\ref{Frad}) practically coincides with the model (\ref{Ftt}), if the parameter $\sigma = 0.48$ GeV$^2$. Note, that these logarithmically enhanced models are not able to describe the $\eta,\eta'$ form factors.

\begin{table*}[htb]
\caption[Results]{Values of the branchings $R\left(  P\rightarrow l^{+}l^{-}\right)=\Gamma\left(  P\rightarrow l^{+}l^{-}\right)/\Gamma_{tot}$ obtained in our approach and compared with the available
experimental results. }%
\label{table2}%\renewcommand{\tabcolsep}{2pc} % enlarge column spacing
\renewcommand{\arraystretch}{1.0} % enlarge line spacing
\begin{tabular}[c]{|c|c|c|c|c|c|}\hline
$R$ & Unitary  & CLEO+BABAR & CLEO+BABAR & With mass & Experiment\\
& bound &  bound & +OPE & corrections  & \\\hline $R\left(  \pi^{0}\rightarrow e^{+}e^{-}\right)  \times10^{8}$ & $\geq4.69$ & $\geq5.85\pm0.03$ & $6.23\pm0.12$ &
$6.26$ & $7.49\pm0.38$ \cite{Abouzaid:2006kk}\\\hline $R\left(  \eta\rightarrow\mu^{+}\mu^{-}\right)  \times10^{6}$ & $\geq4.36$ &
$\leq6.60\pm0.12$ & $5.35\pm0.27$ & $4.76$ & $5.8\pm0.8$ \cite{Amsler:2008zzb,Abegg:1994wx}\\\hline $R\left(  \eta\rightarrow e^{+}e^{-}\right)
\times10^{9}$ & $\geq1.78$ & $\geq4.27\pm0.02$ & $4.53\pm0.09$ & $5.19$ & $\leq2.7\cdot10^{4}$ \cite{Berlowski:2008zz}\\\hline $R\left(
\eta^{\prime}\rightarrow\mu^{+}\mu^{-}\right)  \times10^{7}$ & $\geq1.35$ & $\leq1.44\pm0.01$ & $1.364\pm0.010$ & $1.24$ & \\\hline $R\left(
\eta^{\prime}\rightarrow e^{+}e^{-}\right)  \times10^{10}$ & $\geq0.36$ & $\geq1.121\pm0.004$ & $1.182\pm0.014$ & $1.83$ & \\\hline
\end{tabular}
\end{table*}

However, these approaches still contain some unanswered questions. First of all, the relation of the parameter $\sigma$ in (\ref{Frad}) (or $m$ in \cite{Polyakov:2009je}) to the fundamental QCD parameters is unclear. Secondly, the origin of the flat distribution amplitude is not well justified. Most of the QCD sum rule and instanton model calculations lead to endpoint suppressed amplitudes (see, e.g. \cite{Bakulev:2002uc,Dorokhov:2002iu}). This is a simple consequence of the nonlocal structure of the QCD vacuum. The pion-to-quarks vertex has its own hadronic form factor with the characteristic scale of the vacuum nonlocality (the instanton size). It leads to suppression, if realistic values of parameters are used. Only under the assumption that vacuum nonlocalities disappear, a flat pion distribution amplitude is obtained. Note, that the flatness is a natural property of the photon distribution amplitude, because the photon has no intrinsic hadronic form factor \cite{Dorokhov:2006qm}.

The possible origin of the difference of the asymptotic behavior of the pion and $\eta,\eta'$ meson form factors is the flavor composition of these mesons  \cite{Dorokhov:2009zx}. The pion consists of almost massless $u,d$ quarks, while the $\eta,\eta'$ mesons include also an $s$ quark. The $s$ quark with mass $m_s$ of order $\Lambda_{QCD}$ may be considered as a heavy one. Recently, a similar behavior to that predicted by (\ref{Fcleo}) was found for the $\gamma\gamma^*\to\eta_c$ transition form factor, measured by the BABAR collaboration for the range $Q^2=2-50$ GeV$^2$ \cite{Druzhinin:2009gq}. The corresponding fitted mass parameter is $\Lambda_{\eta_c}=2.92\pm16$ GeV.

Let us consider the possible influence of the BABAR data on the rare leptonic decays of the light pseudoscalar mesons. Remind, that the imaginary part of the amplitude of these decays is fixed unambiguously by the two-photon decay widths of these mesons. Neglecting the real part of the amplitude, the so called unitary bound is obtained\footnote{In general the unitary bound related to the two-photon intermediate state is violated in the case of $\eta'$ meson where new thresholds appear.}. In particular, one has
\begin{eqnarray}
&&R\left(  \pi^{0}\rightarrow e^{+}e^{-}\right)\label{UnitB}  \\
&&\geq R^{\mathrm{unitary}}\left(  \pi^{0}\rightarrow e^{+}e^{-}\right)  =4.69\cdot10^{-8},\nonumber
\end{eqnarray}
which is $7.3\sigma$ below the branching $R^{\mathrm{KTeV}}\left(  \pi^{0}\rightarrow e^{+}e^{-}\right)  =(7.49\pm0.38)\cdot10^{-8}$  obtained by the KTeV collaboration \cite{Abouzaid:2006kk}.

The structure of the real part of the amplitude was considered in detail in \cite{Dorokhov:2007bd,Dorokhov:2008cd,Dorokhov:2009xs}. It consists of the logarithmic (model independent) terms $ln(m^2_l/M_P^2)$ and $ln(m^2_l/\Lambda^2)$, a constant term related to the inverse moment of the pion transition form factor in symmetric kinematics
\begin{equation}
\mathcal{A}\left(  q^{2}=0\right)  =\frac{3}{2}\ln\left(  \frac{m^2_{e}}{\mu^2}\right)
+\chi_{P}\left(  \mu\right)  ,\label{RM0}%
\end{equation}
with
\begin{eqnarray}
&&\chi_{P}\left(  \mu\right)  =-\frac{5}{4}
-\frac{3}
{2}\left[  \int_{0}^{\mu^{2}}dt\frac{F_{\pi\gamma^{\ast}\gamma^{\ast}}\left(
t,t\right)  -1}{t}\right.\nonumber\\&&
\left.+\int_{\mu^{2}}^{\infty}dt\frac{F_{\pi\gamma^{\ast}%
\gamma^{\ast}}\left(  t,t\right)  }{t}\right]  \nonumber,
\end{eqnarray}
and of small mass corrections, $\left(m^2_e/M_P^2\right)^n$, $\left(m^2_e/\Lambda^2\right)^n$, $\left(M_P^2/\Lambda^2\right)^n$, where $\Lambda\approx M_\rho$ is characteristic parameter of the form factor. As was explained in \cite{Dorokhov:2007bd}, the data on the transition form factors $F_{P\gamma\gamma^*}(Q^2,0)$ provide improved lower bounds for electronic decay modes, because in this case the logarithmic terms in the amplitude dominate over the constant term (see Table 1). On the contrary, for the muonic decay modes of the $\eta$ and $\eta'$ mesons, the logarithmic and constant terms are comparable, and by using the data on $F_{P\gamma\gamma^*}(Q^2,0)$ one gets upper bounds for the branchings (Table 1).  The analysis of the CELLO and CLEO data on the pion transition form factor leads to an improved bound
\begin{eqnarray}
&&R\left(  \pi^{0}\rightarrow e^{+}e^{-}\right)  \label{CleoB}\\
&&\geq R^{\mathrm{CLEO}}\left(  \pi^{0}\rightarrow e^{+}e^{-}\right)  \nonumber\\
&&=(5.85\pm0.03)\cdot10^{-8},\nonumber
\end{eqnarray}
which is $4.3\sigma$ below the KTeV result.

Let us check how sensitive these improved bounds are to the recent BABAR data at large momentum transfer. As seen from Figs. 1-3, there are two main changes after the appearance of the BABAR data, compared with the CLEO parametrization. Firstly, the tail of the pion form factor has unexpected asymptotics, and, secondly, $\Lambda_\eta\approx 866$ MeV is closer to the corresponding $\eta'$ parameter. In order to estimate the influence on the pion decay, we choose  the parametrization suggested in \cite{Knecht:2001xc}
\begin{eqnarray}
 &&F_{\pi\gamma\gamma^*}(Q^2,0)=\frac{f_\pi}{3M^2_VM^2_{V'}} \nonumber
\\ && \cdot\frac{h_1Q^4+h_5Q^2+h_7}{(Q^2+M^2_V)(Q^2+M^2_{V'})},
\label{Fnyf}
\end{eqnarray}
where $M_V=769$ MeV, $M_{V'}=1465$ MeV and the parameter $h_7=14.153$ is fixed by the anomaly (\ref{F0exp}). The best fit to the CELLO and CLEO data is given by $h_5=6.93$ and $h_1=0$ \cite{Knecht:2001xc}. The best fit, including the BABAR \cite{:2009mc} data corresponds to \cite{Nyffeler:2009uw} $h_5=6.51$ and switching on the small coefficient $h_1=0.17$, responsible for the asymptotics observed by BABAR (dashed line in Fig. 1)\footnote{However, the parametrization (\ref{Fnyf}) can not be considered as physical one and used for an extrapolation to higher $Q^2$, because it contradicts the Terazawa-West inequality \cite{Terazawa:1973hk,West:1973gd} $F_{\pi\gamma\gamma^*}(Q^2,0)\leq 1/Q$ following from unitarity.}. Comparing the constant (\ref{RM0}), calculated with one (\ref{Fcleo}) or another (\ref{Fnyf}) parameterization of the data, with the integral in (\ref{RM0}), taken in the region from $0$ to $40$ GeV$^2$, we find only small discrepancy and thus no changes for predictions of the $\pi^-\to e^+e^-$ decay (Table 1). A similar conclusion was found for the influence of the BABAR data \cite{:2009mc} on the hadronic light-by-light scattering contribution to the muon anomalous moment \cite{Nyffeler:2009uw}.

The new scale $\Lambda_\eta$ for the  $\eta$ meson form factor, obtained from the inclusion of the BABAR data \cite{Aubert:2006cy}, slightly changes the numbers in the second and third lines of the Table. The fifth column of Table contains the predictions when the mass corrections to the amplitude are taken into account \cite{Dorokhov:2009xs}. These corrections have some influence on the predictions for the $\eta$ and $\eta'$ mesons decays. Thus, it is clear, that more precise data at low energy would lead to stronger restrictions on the leptonic decays widths of the light pseudoscalar mesons.

Further independent experiments for $\pi^0\to e^+e^-$ at WASAatCOSY \cite{Kupsc:2008zz} and for $\eta(\eta')\to l^+l^-$ KLOE \cite{Bloise:2008zz} and BES III \cite{Li:2009jd} and other facilities will be crucial for resolution of the problem with the rare leptonic decays of light pseudoscalr mesons. It is also important to confirm the theoretical basis for a maximally model independent prediction of the branchings (see Table) by getting more precise data on the pion transition form factor in asymmetric as well as in symmetric kinematics in wider region of momentum transfer. Such data are expected soon from the BABAR, BELLE (at large momentum transfer) and KEDR (at small momentum transfer) collaborations.

There are quite few attempts in the literature,  to explain the excess of the experimental data on the $\pi^{0}\rightarrow e^{+}e^{-}$ decay over the SM prediction, as a manifestation of physics beyond the SM. In Ref. \cite{Kahn:2007ru}, it was shown that this excess could be explained within the currently popular model of light dark matter involving a low mass ($\sim10$ MeV) vector bosons $U_{\mu}$, which presumably couple to the axial-vector currents of quarks and leptons. Another possibility was proposed in Ref. \cite{Chang:2008np,McKeen:2008gd}, interpreting the same experimental effect as the contribution of the light CP-odd Higgs boson appearing in the next-to-minimal supersymmetric SM.

The author thanks S.B. Gerasimov, V.P. Druzhinin, N.I. Kochelev, E.A. Kuraev, S.V. Mikhailov, A.A. Pivovarov, A.V. Radyushkin for discussions on the interpretation of the high momentum transfer data for the pseudoscalar meson transition form factors.

\end{document}